# Hot-carrier relaxation in photoinjected ZnSe

Clóves G. Rodrigues

*Núcleo de Pesquisa em Física, Departamento de Matemática e Física, Universidade Católica de Goiás, CP 86, 74605-010, Goiânia, Goiás, Brazil*

cloves@pucgoias.edu.br

**Abstract**

A theoretical investigation of the excess energy dissipation of highly excited photoinjected carriers in zinc selenide (ZnSe) is presented. The calculations are performed by solving numerically coupled quantum transport equations for the carriers and the optical phonons in order to derive the evolution of their nonequilibrium temperatures, dubbed quasitemperatures (or nonequilibrium temperatures). It is shown that the carrier energy dissipation occurs in a picosecond time scale.

## 1. Introduction

The zinc selenide (ZnSe) is a wide-gap II–VI zincblende semiconductor. The ZnSe is an attractive semiconductor for various optoelectronic devices, for instance laser diodes [1–3].

We consider here the carrier relaxation kinetics in a highly excited photoinjected double plasma generated in ZnSe. We recall that a HEPS consists of electrons and holes as mobile carriers (created by an intense ultrashort laser pulse), which are moving in the background of lattice vibrations. These photoexcited carriers with a concentration $n$ are initially narrowly distributed around the energy levels in conduction and valence bands separated by the photon energy $\hbar\Omega$. After this initial stage they proceed to redistribute their excess energy as a result of strong Coulomb interaction. At the same time relaxation processes occur transferring the excess energy to the lattice and also to the media surrounding the active volume of the sample.

On the other hand, the concentration diminishes in time via the processes of recombination and ambipolar diffusion [4]. Assuming the modelling used in [5], we take the nonequilibrium carrier distributions in photoexcited ZnSe as described by distribution functions characterized by quasitemperatures well above that of the lattice. The energy relaxation of the carriers follows mainly through scattering processes with phonons (dominated by LO phonons) [6].

We study in this work the hot carrier dynamics using coupled quantum transport equations for the carrier quasitemperature (after Coulomb thermalization) and the acoustic and longitudinal optical phonon temperatures to have a picture of their excess energy dissipation.

## 2. Model and method

We deal theoretically with the hot-carrier dynamics in ZnSe, which is basically a problem involving a nonlinear nonequilibrium kinetic, resorting to a theory built within the framework of a particular nonequilibrium ensemble formalism, the so-called nonequilibrium statistical operator method (NESOM) [7] and Zubarev's approach is used [8]. Moreover, we introduce the Markovian approximation to the theory [9].

To proceed further it is necessary to characterize the nonequilibrium macroscopic (i.e. nonequilibrium thermodynamic) state of the system, the so-called kinetic stage appropriate for its description in the given experimental conditions (as discussed in [10]) mainly considering the extension of the pulse of the exciting laser source and the resolution time of the detector. First, we notice that a single-quasiparticle description (allowing to introduce a band structure for the electron energies and the use of the random phase approximation) can follow after a very initial transient time of the order of a period of a plasma wave, which is typically in the tenfold femtosecond scale. The next kinetic stage corresponds to a description of the carriers in terms of distribution functions in single-particle band energy states. Next, under the action of Coulomb interaction and carrier-phonon collisions there follows an internal thermalization of the carriers whose nonequilibrium thermodynamic state can be characterized by the time-dependent density $n(t)$ and quasitemperature $T^*_{(t)}$ [10,11]. In most cases a third kinetic stage is characterized by the mutual thermalization of carriers and optical phonons, followed by the attainment of final equilibrium with the acoustical phonons and the external thermal reservoir.

Table 1
Parameters of ZnSe

| Parameter | Value |
|---|---|
| Electron effective mass[a] | $m^*_e = 0.15\text{m}_0$ |
| Hole effective mass[a] | $m^*_h = 0.67\text{m}_0$ |
| LO phonon energy[a] | $\hbar\omega_{\text{LO}} = 31.5\,\text{meV}$ |
| Static dielectric constant[a] | $\varepsilon_0 = 7.6$ |
| Optical dielectric constant[a] | $\varepsilon_\infty = 5.4$ |
| Optical relaxation time[a] | $\tau_{op} = 4.5\,\text{ps}$ |
| Lattice parameter[b] | $a = 5.65\,\text{Å}$ |
| Band gap energy[c] | $E_g = 2.69\,\text{eV}$ |

[a]Ref. [13].
[b]Ref. [14].
[c]Ref. [15].

## 3. Results and discussion

We consider here an ultrashort pumping laser pulse producing in ZnSe an initial concentration of carriers $10^{18}\,\text{cm}^{-3}$ and excess energy of 1.2 eV to avoid intervalley scattering. We began to consider the second kinetic stage, i.e. we assume that Coulomb thermalization among the carriers has already occurred, and we take for them an initial carrier quasitemperature of 4640 K (corresponding to the excess energy per carrier of 1.2 eV). The temperature of the lattice is 300 K, and we consider the case a very good thermal contact keeps the acoustic phonons constant at the reservoir temperature. We derive the equations of evolution for the carrier and LO phonon quasitemperatures through the NESOM within the Markovian approximation. We omit the details of the calculations which follow the same scheme as those in [12] for GaAs. In GaAs the investigation of carrier relaxation kinetics can be affected by intervalley scattering because the satellite $L$ valley is only 300 meV above the $\Gamma$ valley conduction minimum, however, in the ZnSe this complication is reduced, since the $\Gamma$–$L$ valley energy splitting is approximately 1 eV. Table 1 presents the ZnSe parameters used in the calculations.

In Fig. 1 we show the evolution of the carrier quasitemperature for the times involved (picosecond scale) when the concentration $n(t)$ is practically unaltered.

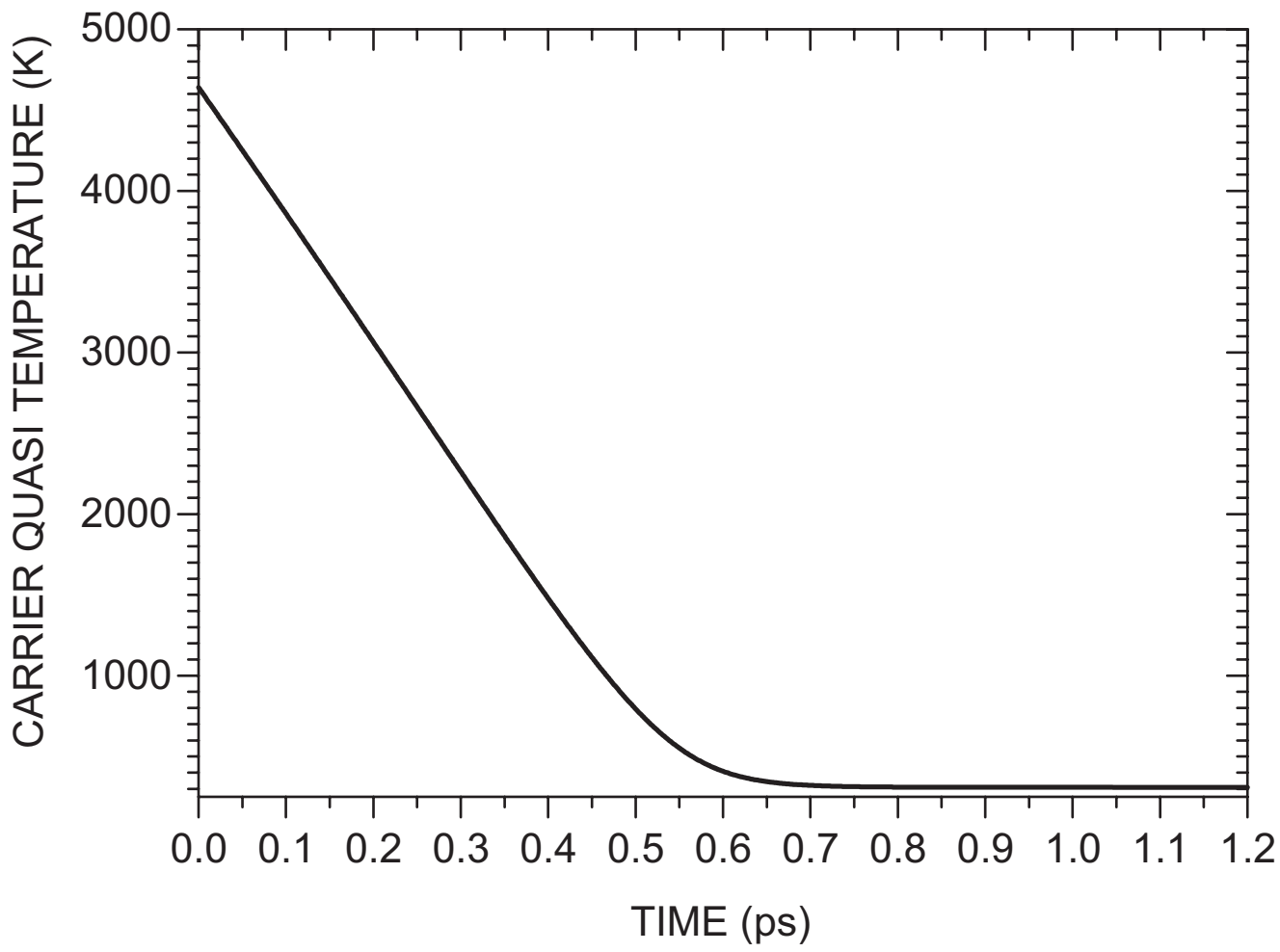

Fig. 1. Time evolution of the carrier quasitemperature in ZnSe. The initial photoinjected carrier density and temperature are $10^{18}\,\text{cm}^{-3}$ and 4640 K, respectively. The temperature of reservoir is maintained at 300 K.

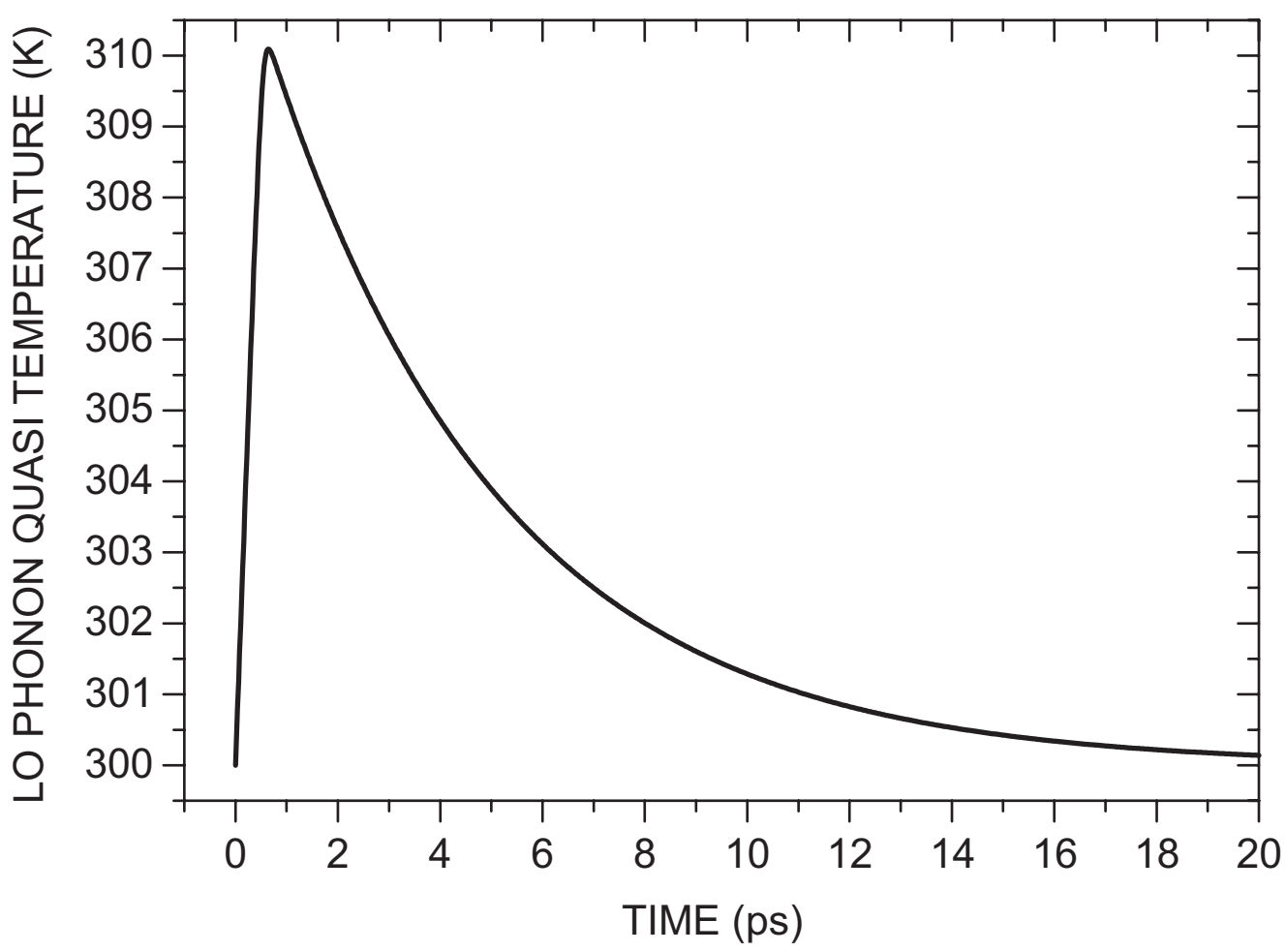

Fig. 2. Time evolution of the LO phonon quasitemperature in ZnSe. The initial LO phonon quasitemperature is 300 K. The temperature of reservoir is maintained at 300 K.

Inspection of Fig. 1 tells us that occurs a very rapid relaxation of the carrier excess energy to the lattice (with a characteristic time of roughly 1 ps), together with a final thermalization and equilibrium in, say, 1.2 ps occur. In comparison with gallium nitride (GaN) (see Ref. [16]), the energy dissipation is shown to be faster in GaN than in ZnSe. This is clearly a consequence of III-nitirides (which are strong polar semiconductors with intense Fröhlich interaction between carriers and LO phonons), particularly the GaN, being more strongly polar than the ZnSe, with a Fröhlich coupling constant more intense in GaN than in ZnSe. We recall that Fig. 1 was obtained starting at the third kinetic stage after assuming Coulomb thermalization of the carriers.

Fig. 2 shows the evolution of the LO phonon quasitemperature in the range of 0–20 ps. We can see that the LO phonons are warmed up and acquire, along approximately the first 0.6 ps, quasitemperatures up to 310 K. These excited LO phonons relax their excess energy to thermal bath follows in a 20 ps timescale.

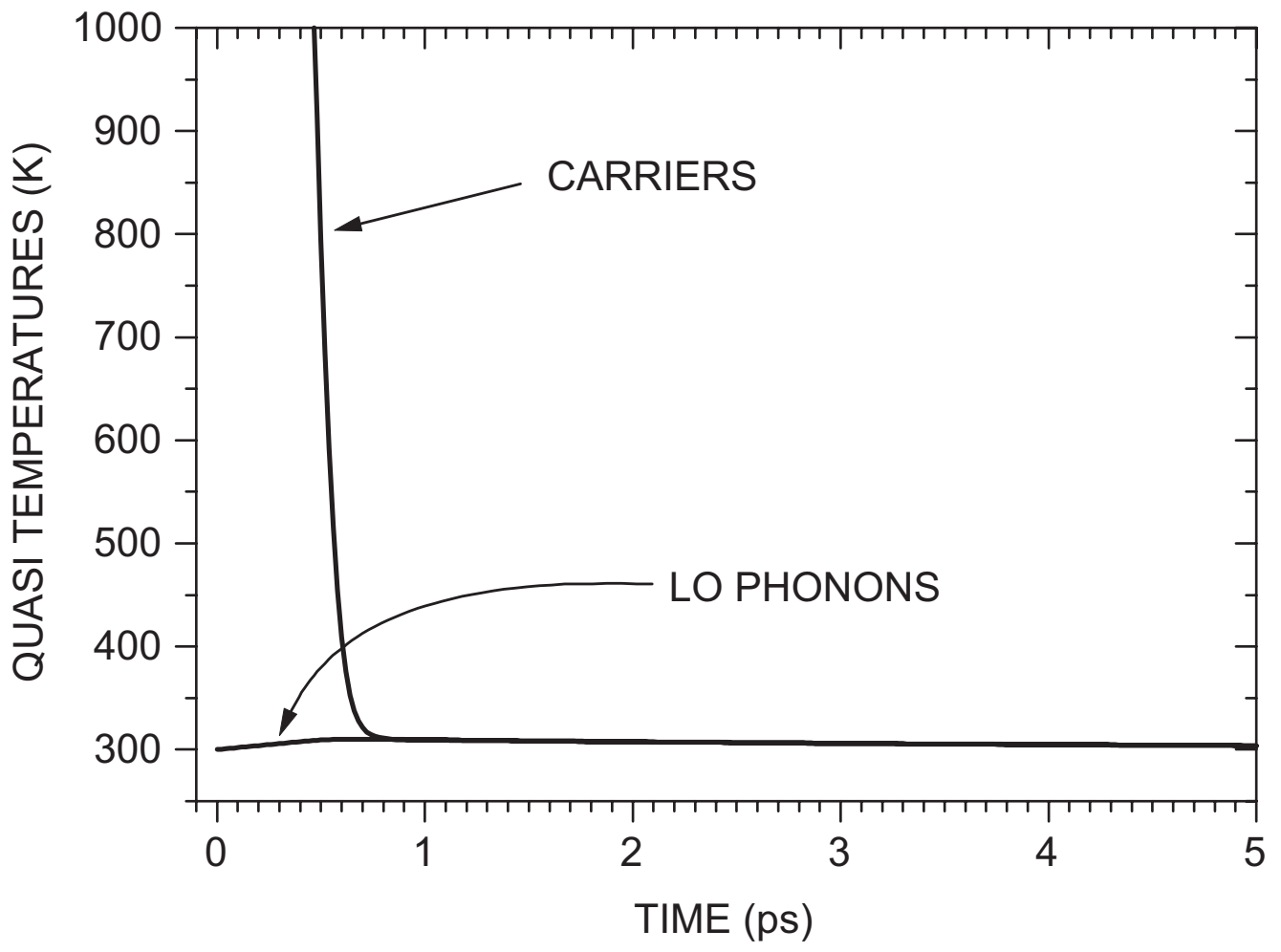

Fig. 3. Time evolution of the carriers and LO phonon quasitemperatures.

Fig. 3 shows the evolution of the carrier and LO phonon quasitemperatures in the range of 0–5 ps and 300–1000 K. We note that the mutual thermalization of carriers and LO phonons follows in, roughly, 1 ps, accompanied soon after by a final thermal equilibrium with the thermal reservoir.

We reminded that in the timescales involved (up to approximately 20 ps) the change in concentration due to recombination of electrons and holes, and ambipolar diffusion out of the active volume of the sample is negligible [17].

## 4. Final comments

We present here a study concerning the photoexcited carrier dynamics in zincblende ZnSe. Photoexcited excess carrier energy relaxation in semiconductors is one important subject that is only recently receiving attention, as a result of the relevance for understanding how the excess carrier energy dissipation influences the working of devices [5,6,18,19].